\documentclass{iaus}
\usepackage{graphicx}

\title[Kinematic substructures in the Coma Cluster Core] 
{Kinematic Substructures in the Coma Cluster Core as traced by
  Intracluster Planetary Nebulae}

\author[M.~Arnaboldi etal.]   
{Magda Arnaboldi$^1$, Ortwin Gerhard$^2$, Kenneth C. Freeman$^3$,\break 
Nobunari Kashikawa$^4$, Sadanori Okamura$^5$, Naoki Yasuda$^6$ }

\affiliation{$^1$ESO \& INAF - Obs Turin,
$^2$MPE - Garching, $^3$ RSAA Canberra, $^4$ NAOJ Tokyo, $^5$ Dpt. of
  Astron. Tokyo, $^6$ Inst. Cosm. Ray Res. Tokyo\break email: marnabol@eso.org}

\pubyear{2006}
\volume{234}  
\pagerange{1--4}
\date{7.~April 2006} 
\jname{Planetary Nebulae in our Galaxy and Beyond}
\editors{M. Barlow, R.H. Mendez, eds.}
\begin{document}

\maketitle

\begin{abstract}
The Coma cluster is the richest and most compact of the nearby
clusters, yet there is growing evidence that its formation is still
on-going. A sensitive probe of this evolution is the dynamics of
intracluster stars, which are unbound from galaxies while the cluster
forms, according to cosmological simulations. With a new multi-slit
imaging spectroscopy technique pioneered at the 8.2 m Subaru telescope
and FOCAS, we can now detect and measure the line-of-sight velocities
of the intracluster planetary nebulae which are associated with the
diffuse stellar population of stars, at 100 Mpc distance. We detect
significant velocity substructures within a 6 arcmin diameter field,
centred on the Coma X-ray cluster emission. One substructure is
present at $\sim 5000$ kms$^{-1}$, probably from infall of a galaxy
group, while the main intracluster stellar component moves at $\sim
6500$ kms$^{-1}$. Hence the ICPNs associated with the diffuse light at
the position of the MSIS field are not bound to the nearby cD galaxy
NGC 4874, whose radial velocity is $\sim 700$ kms$^{-1}$ greater.

\end{abstract}

\firstsection 
\section{Introduction}
Diffuse intracluster light (ICL) has now been observed in nearby
(Feldmeier et al. 2004, Mihos et al. 2005) and in intermediate
redshift clusters (Zibetti et al. 2005, Krick et al. 2006). Recent
studies show that the ICL contains of order 10\% of the mass in stars
overall (Aguerri et al. 2005, Zibetti et al. 2005), but in the cores
of dense and rich clusters like Coma, the local ICL fraction can be as
high as 50\% (Bernstein et al. 1995).

Intracluster planetary nebulae (ICPNs) are presently the only tracers
which allow us to measure the {\it kinematics} and {\it dynamics} of
the ICL, using their [OIII] 5007\AA\ emission for identification and
radial velocity measurements. By determining the projected phase space
distribution of ICPNs we can constrain the dynamical age of the ICL
component, how and when this light originated. Owing to the small
fluxes of distant ICPNs this was so far only possible for the nearby
Virgo cluster (Arnaboldi et al. 2004). 

To study the effect of a dense environment on galaxy evolution, we
have recently carried out a pilot survey in the Coma cluster (A1656),
using a multi-slit imaging spectroscopy (MSIS) technique with the
FOCAS spectrograph and the Subaru telescope (Gerhard et al. 2005).
First detections and line-of-sight velocities for one mask
configuration were presented in that paper. Here we discuss the result
from the data reduction and analysis of the whole dataset. We present
the spatial and velocity distribution of 40 detected PN candidates and
compare with those of Coma cluster galaxies from the most recent
compilation by Adami et al. (2005).

\section{Observations}\label{MSISobs}

Observations of a field centered on $\alpha$(J2000) 12:59:41.784;
$\delta$(J2000) 27:53:25.388 were carried out with the Faint Object
Camera and Spectrograph (FOCAS) at the 8.2 m Subaru telescope on April
21-23, 2004. This field is near the X-ray centroid of the Coma cluster
and is essentially coincident with the field observed by Bernstein et
al. (1995).  The instrument was used with a mask of parallel multiple
slits (shown in O.~Gerhard's review at this conference), and with a
narrow band filter centered on the wavelength of the redshifted [OIII]
5007\AA\ line at the mean velocity of the Coma cluster. We observed
three MSIS masks to obtain spectra of all PNs that happen to lie
behind the slits.

The spatial resolution of FOCAS is $0''.1$ pixel$^{-1}$, so the $6'$
diameter (equivalent to 174 kpc in Coma) of the circular field of view
(FOV) corresponds to 3600 pixels. We used grating 300B, which gives a
measured dispersion of 1.41 \AA\ pixel$^{-1}$ on the two FOCAS CCD
chips. The effective spectral resolution is $\simeq 7.3$ \AA, or $440$
kms$^{-1}$.  We used the N512 filter with an FWHM of 60 \AA, centered
at $\lambda_c = 5121 \AA$. The FWHM includes only $\pm 1.6 \times
\sigma$ the galaxy velocity dispersion in the Coma core. In the
current instrument configuration the dispersion axis is the $y-$axis,
while the spatial direction is along the $x-$axis. For a detailed
description of the technique and signal-to-noise calculations we refer
to Gerhard et al. (2005).

\subsection{Data reduction}\label{MSISdared}

The complete data reduction of the three observed MSIS masks was
carried out in IRAF and included the following standard steps:
correction for bias/flatfield, frames alignment and co-addition.  We
had 13 frames in the first two masks respectively, and 6 frames for
the third mask; the final three coadded MSIS images for the 3 masks
were then free of cosmic rays.  Flux calibration was done using the
spectrophotometric standard star BD $+30^\circ26'42''$.

The median averaged MSIS frame for each night was inspected for the
presence of emission line objects. Monochromatic, point-like emitters
appear as elongated ellipses on the CCD, with a width which depends on
the binning adopted for the $x-$coordinate and seeing ($\sim 3$ pix
for 1$^{st}$ and 2$^{nd}$ night, 4 pix for the $3^{rd}$ night) and a
height of 5 pixels in the wavelength direction.

In these scientific frames we looked for three kinds of emission line
objects:
\begin{itemize}
\item unresolved (both in wavelength and in space) emission line
objects with no continuum, which are our ICPN candidates;
\item continuum point-like sources with unresolved/resolved line
emissions. These are most likely background galaxies;
\item unresolved line emissions associated with the extended continuum
halo of any of the Coma galaxies in the field - these are compatible
with being PNe associated with the stellar population emitting the
continuum light. In this case, a check on the measured LOS velocity is
required to confirm the PN candidate -- galaxy association.
\end{itemize}
2D slit spectra which contained one of these objects were then
extracted, rectified and wavelength calibrated.

A detailed description of the data reduction, including astrometry and
positions of the emission line candidates, is described in Arnaboldi
et al., in preparation.

\section{Spatial and velocity distribution of ICPNe in the MSIS 
field}\label{posvel}

\begin{figure}
 \includegraphics[height=3in,width=5in]{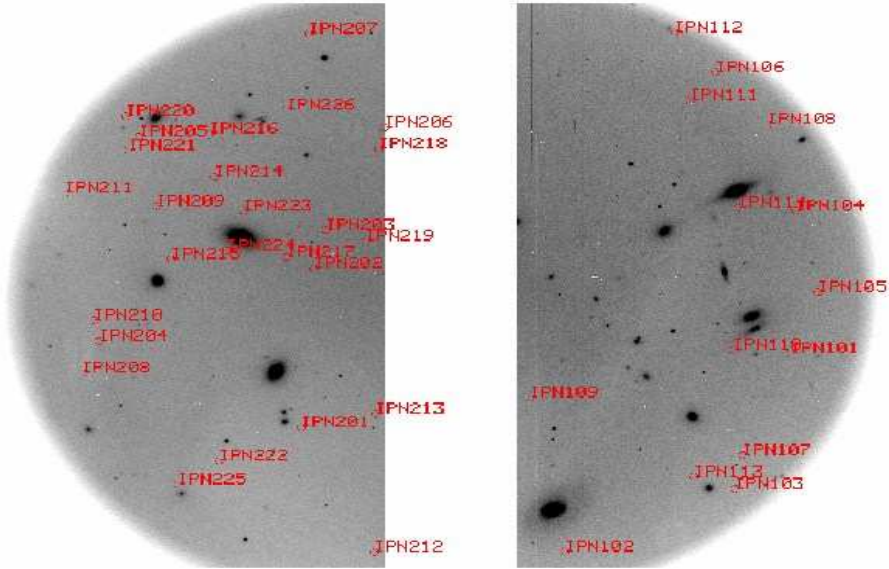}
\caption{Distribution of detected ICPNe in the FOCAS FOV near the
X-ray peak in the Coma cluster. The circular FOV is imaged onto two
CCDs.  The lack of PNe in a fraction of the right-hand side image of
chip \#1 is due to an area on the detector with hot columns and pixels
which cause the instrument noise to increase significantly.
}\label{fig:pos}
\end{figure}

\begin{figure}
\setlength{\unitlength}{0.01\textwidth}
\begin{picture}(50,62)
 \put(0,0)  {\includegraphics[width=48\unitlength]{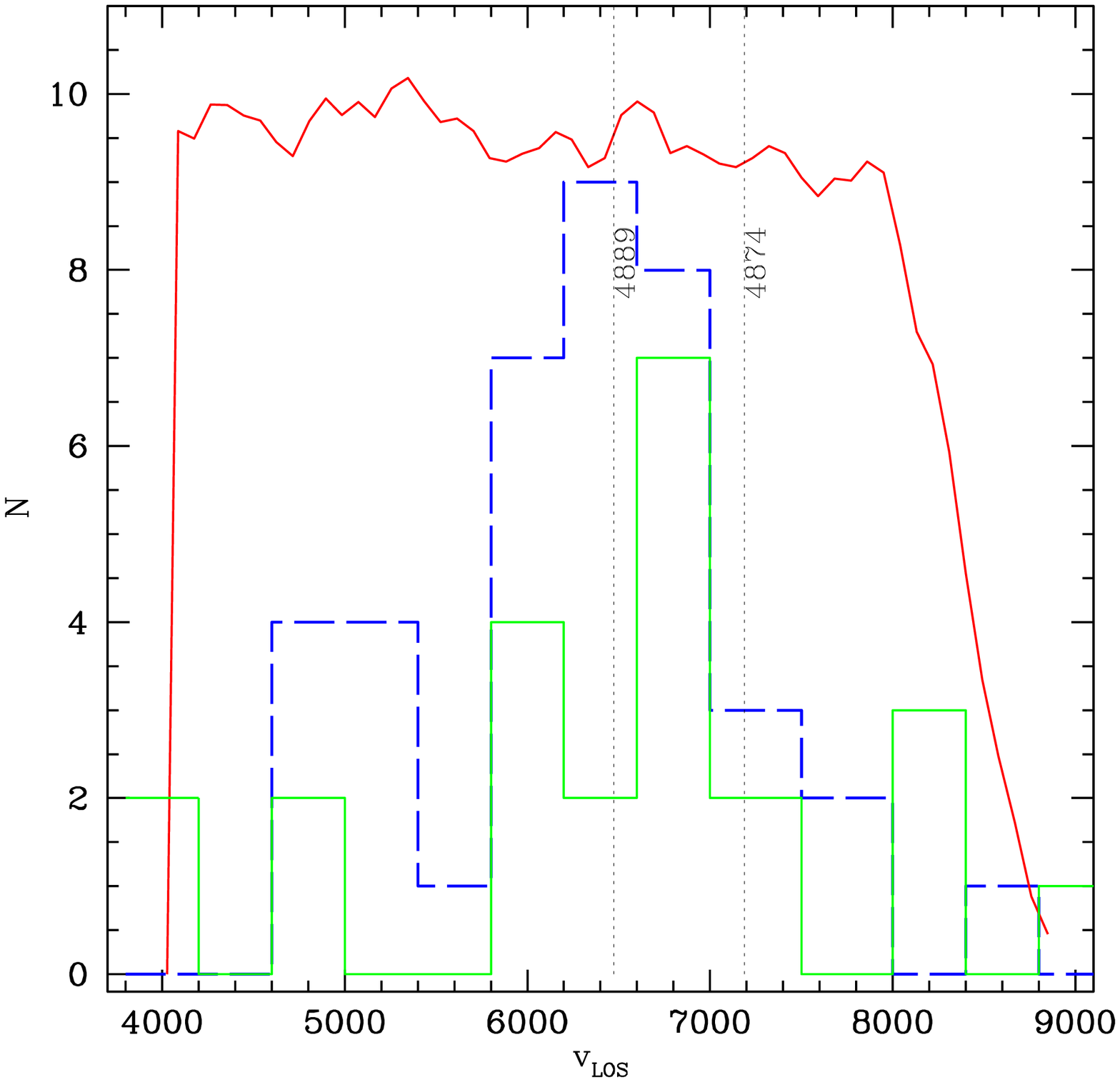}}
 \put(50,0) {\includegraphics[width=48\unitlength]{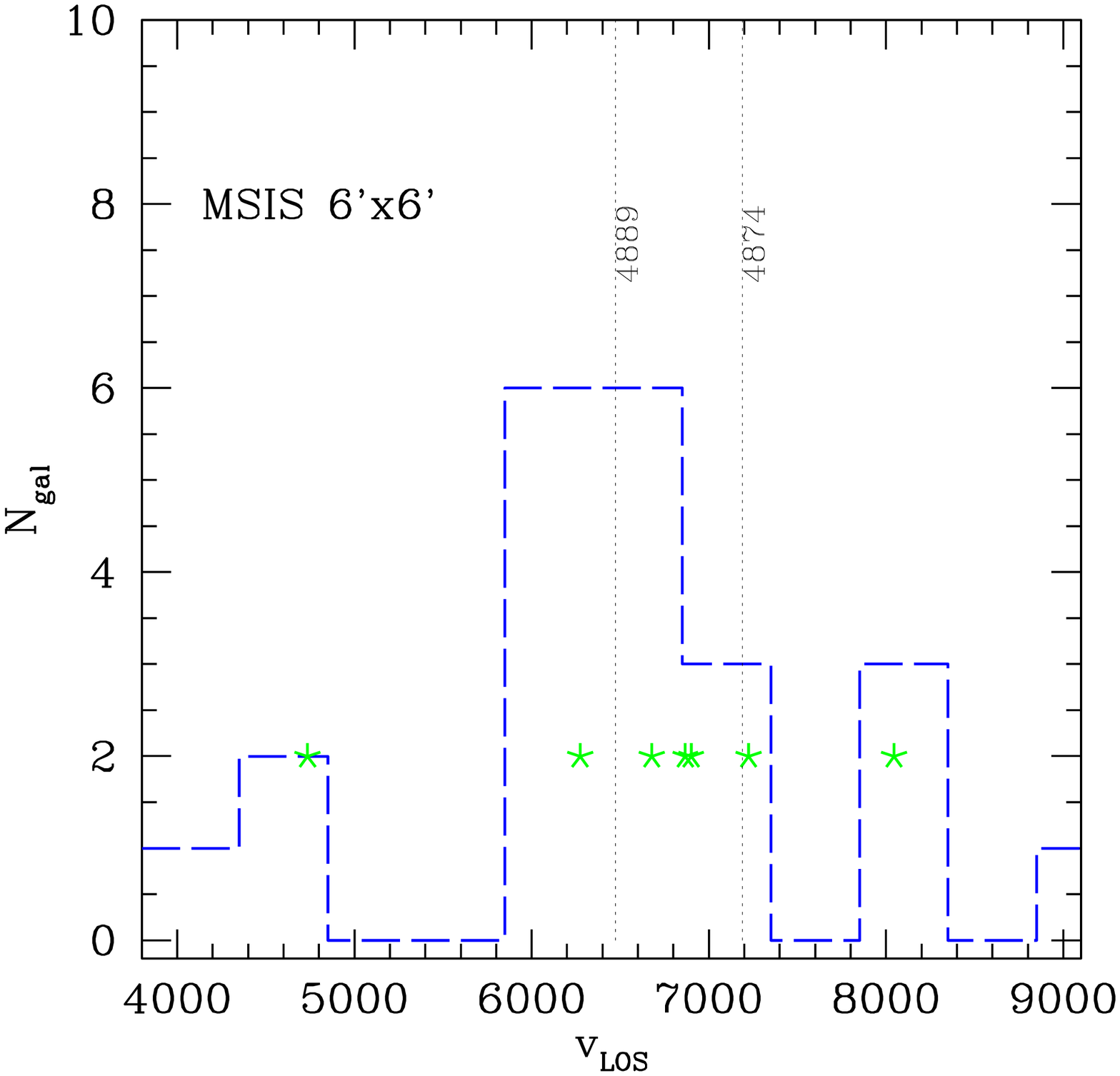}}
\end{picture}
\caption{{\bf (a), left:} Histogram of LOS velocities of ICPNs in the
Coma field (blue dashed lines) and of Coma galaxies in the field (full
green lines).  The upper red curve shows the spectra of continuum
emitters seen through the grism, narrow band filter, and MSIS slits.
{\bf (b), right:} LOS velocity histogram of all Coma galaxies in a
$6'\times 6'$ square centered on the MSIS field position, which would
fall in the velocity window as indicated by the red line in the left
panel (blue dashed lines). The Coma galaxy redshifts are from Adami et
al. (2005). The peak of the distribution is at $\sim 6500$ kms$^{-1}$
and the distribution is clearly not Gaussian. In both panels, the
redshifts for the two cD galaxies NGC 4889 and NGC 4874 are indicated
by vertical dotted lines; NGC 4874 is clearly displaced from the main
peak of the velocity distribution.
\label{fig:vel}}
\end{figure}

Figure~\ref{fig:pos} shows the spatial distribution of all IPCN
candidates in the MSIS field, and Figure~\ref{fig:vel}a shows the LOS
velocity distribution of the ICPNe associated with the diffuse light
in the Coma cluster core. The LOS velocities of the PN candidates show
a clear association with the Coma cluster. The average velocity of the
distribution is 6315 kms$^{-1}$ and the standard deviation is 867
kms$^{-1}$, but two substructures are also clearly visible, with a
main peak at $v\sim 6500$ kms$^{-1}$ and a secondary peak at $v=5000$
kms$^{-1}$. The main peak in the LOS velocity distribution of PN
candidates is close to the systemic velocity of NGC 4889, $v_{sys} =
6495$ kms$^{-1}$.

Figure~\ref{fig:vel}b shows the $v_{LOS}$ velocity distribution of the
Coma galaxies in a $6'\times 6'$ field centred on the MSIS position.
This has multiple peaks: the largest peak is coincident with the
$v_{sys}=6495$ kms$^{-1}$ of NGC~4889, and two less prominent peaks
are at $v_{LOS} < 5000$ kms$^{-1}$ and $v_{LOS} \sim 8100$ kms$^{-1}$.


The main peak of the ICPN velocity histogram coincides with that of
the Coma galaxy velocities around the MSIS field, and it is centered
on the systemic velocity of NGC~4889 at $v_{sys}=6495$ kms$^{-1}$.
The galaxy NGC~4874, which lies nearer to the MSIS field, the peak of
the X-ray emission, and the center of the larger-scale X-ray emission
(White et al 1983), is at $v_{sys} = 7189$ kms$^{-1}$, displaced by
$\sim 700$ kms$^{-1}$ from the velocity peaks for both the galaxies
and the ICPNs. Hence the ICPNs associated with the diffuse light at
the position of the MSIS field are not bound to the nearby NGC 4874,
but have motions similar to those of the Coma galaxies in this field.

\begin{acknowledgments}
We would like to acknowledge the financial support by the Swiss SNF
and by INAF. We are grateful to the on-site Subaru staff for their
support.  
\end{acknowledgments}

\end{document}